\documentclass{iaus}
\usepackage{graphicx}

\title[ULX nebula MF16] %% give here short title %%
{The Optical Counterpart of an Ultraluminous X-ray Source  NGC6946~ULX-1}

\author[Abolmasov, Fabrika \& Sholukhova]   %% give here short author list %%
{Pavel K. Abolmasov$^1$
  \thanks{Present address: Special Astrophysical Observatory of the Russian AS,
Nizhnij Arkhyz 369167, Russia},
 S. N. Fabrika$^1$ \and O. N. Sholukhova$^1$}

\affiliation{$^1$ Special Astrophysical Observatory of the Russian AS,
Nizhnij Arkhyz 369167, Russia \break email: pasha@sao.ru}
\pubyear{2006}
\volume{238}  %% insert here IAU Symposium No.
\pagerange{001--999}
\date{??? and in revised form ???}
\jname{Black Holes: from Stars to Galaxies -- across the Range of Masses}
\editors{V. Karas \& G. Matt, eds.}

\newcommand{\ergl}{ergs~s$^{-1}$}

\newcommand{\feiii}{[Fe~{\sc iii}]}
\newcommand{\hii}{H~{\sc ii}}

\newcommand{\kms}{$km\,s^{-1}\,$}

\newcommand{\Msun}{$M_\odot$}

\renewcommand{\mag}[1]{^{\rm m}\!\!\!#1\,}

\newcommand{\aj}{\textit{AJ}}
\renewcommand{\aa}{\textit{A\&A}}
\newcommand{\apj}{\textit{ApJ}}
\newcommand{\mnras}{\textit{MNRAS}}
\newcommand{\apjs}{\textit{ApJSS}}

\newcommand{\pasp}{\textit{PASP}}

\begin{document}

\maketitle

\begin{abstract}
We present a study of a peculiar nebula MF16 associated with an Ultraluminous X-ray Source NGC6946 ULX-1. We use integral-field and long-slit spectral data obtained with the 6-m telescope (Russia).
%The nebula is found to be very dense ($n_e \sim 10^3 cm^{-3}$), highly asymmetric and expanding with about 100-200~\kms.
% A velocity gradient was detected in the integral-field data.
% The expansion velocity is similar to that of a Supernova Remnant (SNR) of a comparable size.
The nebula was for a long time considered powered by strong shocks
enhancing both high-excitation and low-excitation lines. However,
kinematical properties point to rather moderate expansion rates ($V_S \sim 100\div 200$\kms).
The total power of the emission-line source exceeds by one or two orders of magnitude the power observed expansion rate can provide, that points towards the existence of an additional source of excitation and ionization.
Using CLOUDY96.01 photoionization code we derive the properties of the photoionizing source.
Its total UV/EUV luminosity must be about $10^{40}$~erg/s.
%Such an ultraviolet luminosity is too high for a standard accretion
%disk around an Intermediate Mass Black Hole, but could be produced by
%a Supercritical Accretion disk like that in SS433. The observed velocity gradient can be understood as an impact of a collimated wind or jets.
\end{abstract}

\firstsection % if your document starts with a section,

\section{Introduction}\label{sec:intro}

%\subsection{Optical Counterparts of ULXs}

%Ultraluminous X-ray Sources (ULXs) is a
%            phonomenologically-distinguished group of compact
%              non-nuclear extragalactic X-ray sources with luminosities higher than $\sim 10^{39}$\ergl. 
%Their physical nature is unclear, two hypotheses are usually applied:
%more massive accretors -- Intermediate Mass Black Holes, IMBHs
%\cite{comil}
%or
%supercritical accretion combined with the anisotropy of the X-ray
%source \cite{fame,?}.
%More than 150 sources are proved to be real ULXs \cite{swartz}, and
%for large part of them optical counterparts were found. 

Quite a large number of Ultraluminous X-ray Sources (ULXs) are associated
with emission-line nebulae (ULX Nebulae, ULXNe), mostly large-scale
bubbles powered by shock waves \cite[(Pakull \& Mirioni, 2003)]{pamir}. However, several exceptions are known like the nebula
associated with HoII~X-1 \cite[(Lehmann \etal, 2005)]{lehmann}, that is clearly a
photoionized \hii\ region. Another well-known example is the nebula
MF16 coincident with the ULX NGC6946~ULX1.

%\subsection{MF16}

The attention to MF16 was first drawn by \cite{BF_94},
who identified the object as a Supernova Remnant (SNR), according to the 
emission-line spectrum with bright collisionally-excited lines. 
It was long considered an unusually luminous SNR, because of its huge 
optical emission-line ($L_{H\alpha} = 1.9\times10^{39}erg\,s^{-1}$, according to \cite{BF_94},
for the tangential size $20\times34pc$)
and X-ray ($L_X = 2.5\times10^{39}erg\,s^{-1}$ in the $0.5-8$keV range,
according to the luminosities given by \cite{RoCo}).

However, it was shown by \cite{RoCo}, that
the spectral, spatial and timing properties of the X-ray source do 
not agree with the suggestion of a bright SNR, but rather suppose
a point source with a typical ``ULX-like'' X-ray spectrum: cool 
Multicolor Disk (MCD) and a Power Law (PL) component. So, apart from the 
physical nature of the object, MF16 should be considered a 
{\it ULX nebula}, one of a new class of objects, described 
by \cite{pamir}. 

\section{Optical Spectroscopy}\label{sec:obs}

All the data were obtained on the SAO 6m telescope, Russia. Two
spectrographs were used: panoramic MultiPupil Fiber Spectrograph MPFS
\cite[(Afanasiev \etal, 2001)]{MPFSdesc} and SCORPIO focal reducer 
\cite[(Afanasiev \& Moiseev, 2005)]{scorpio} in long-slit mode. 
The details of data
reduction processes and analysis technique will be presented in \cite{mf16_main}.
Panoramic spectroscopy has the advantage of providing unbiased flux
estimates. However, SCORPIO results have much higher signal-to-noise
ratio and reveal rich emission-line spectrum of \feiii.
We also confirm the estimates of the total nebula emission-line
luminosities by \cite{bfs}. $H\beta$ line luminosity obtained from our
MPFS data is $L(H\beta) = (7.2\pm0.2)\times10^{37}erg\,s^{-1}$.

Using line ratios for the integral spectrum we estimate the mean
parameters of emitting gas as: $n_e \simeq 500\pm 100 \,cm^{-3}$, $T_e \simeq (1.9\pm0.2) \times 10^4 K$.
Interstellar absorption is estimated as $A_V \sim 1\mag{.}3$, close to
the Galactic value ($A_V^{Gal} = 1\mag{.}14$, according to \cite{schlegel_abs})

We confirm the estimate of the expansion rate obtained by
\cite{dunne}, coming to the conclusion that the expansion velocity is
$V_S \lesssim 200$\kms. In this case the total emission-line
luminosity can be estimated using for example the equations be
\cite{DoSutI}:

$$
 \begin{array}{l}
F_{H\beta} = %F_{H\beta,shock}+F_{H\beta,precursor} = 
  7.44 \times 10^{-6} \left(  \frac{V_s}{100 km\, s^{-1}} \right)^{2.41} 
\times \left( \frac{n_2}{cm^{-3}}\right)
+ \\
\qquad{} 
9.86 \times 10^{-6} \left(  \frac{V_s}{100 km \,s^{-1}} \right)^{2.28}
\times \left( \frac{n_1}{cm^{-3}}\right) \, erg\, cm^{-2} s^{-1}
\end{array}
$$

Here $V_S$ is the shock velocity and $n_1$ the pre-shock hydrogen
density.
If the surface area is known, one can obtain the total luminosity in
$H\beta$ from here. For $V_S = 200km/s$ and $n_1 = 1cm^{-3}$ it
appears to be $L(H\beta) \simeq 1.6 \times 10^{36}$\ergl, that is too
low compared to the observed value. 
So we suggest an additional source of power providing most of the
energy of the optical nebula.

\section{Photoionization Modelling}

We have computed a grid of CLOUDY96.01~\cite[(Ferland \etal, 1998)]{cloudy98} photoionization models
in order to fit MF16 spectrum avoiding shock waves. We have fixed X-ray
spectrum known from {\it Chandra} observations \cite[(Roberts \&
  Colbert, 2003)]{RoCo}, assuming
all the plasma is situated at 10pc from the central point source,
and introduced a
blackbody source with the temperature changing from $10^3$ to $10^6$K and
integral flux densities from 0.01 to 100 $erg\,cm^{-2}\,s^{-1}$.

The best fit parameters are $\lg T(K) = 5.15\pm 0.05, F = 0.6\pm 0.1 erg\,cm^{-2}\,s^{-1}$,
that suggests quite a luminous ultraviolet source: $L_{UV} =
(7.5\pm0.5) \times 10^{39} erg\,s^{-1}$.The UV source is more than 100
times brighter then what can be predicted by
extrapolating the thermal component of the 
best-fit model for X-ray data \cite[(Roberts \& Colbert, 2003)]{RoCo}.

\section{Ultraluminous UV sources?}

At least for one source we have indications that its X-ray spectrum
extends in the EUV region. It is interesting to analyse the
implications in the frameworks of two most popular hypotheses
explaining the ULX phenomenon.

For the standard disk of \cite{ss73} the inner temperature scales as:

$$
T_{in} \simeq 1~keV\,  \left(\frac{M}{M_\odot}\right)^{-1/4}  \left(\frac{\dot{M}}{\dot{M}_{cr}}\right)^{1/4}
$$

\begin{figure}
\includegraphics[angle=90, width=\textwidth, clip=0]{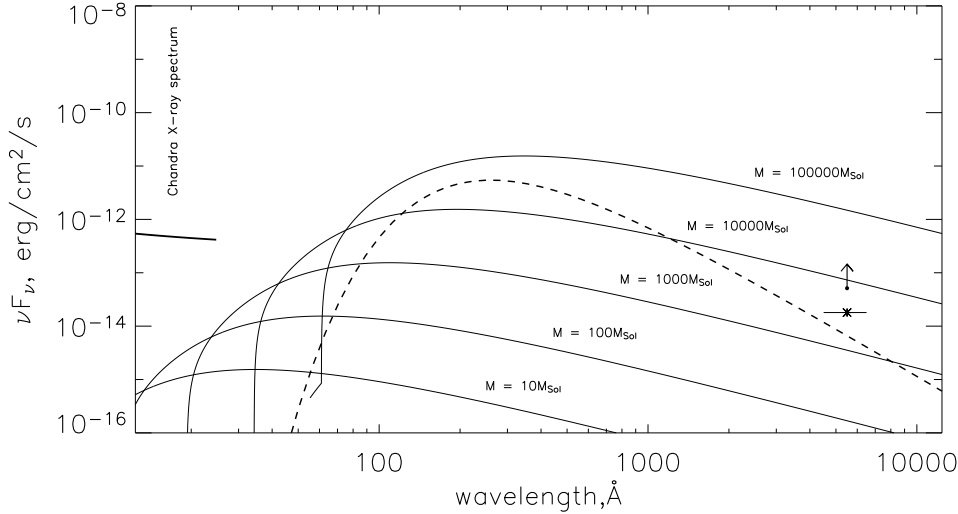}
\caption{NGC6946~ULX1 SED reconstruction. Optical source $d$ \cite[(Blair \etal, 2000)]{bfs} is
shown by an asterisk, and the upward arrow above indicates the
unabsorbed optical luminosity: it is the lower estimate
because only Galactic absorption was taken into account, $A_V =
1\mag{.}14$ according to \cite{schlegel_abs}. Dashed line represents
the best-fit blackbody from our CLOUDY fitting. Thin solid lines are
MCD models for accreting IMBHs with infinite outer disk
radii. Mass accretion rate was set everywhere to $0.01 \dot{M}_{cr}$. 
}\label{fig:seds}
\end{figure}

In Fig.~\ref{fig:seds} we present the reconstructed Spectral Energy
Distribution of NGC6946~ULX-1 including optical identification
by \cite{bfs} and the best-fit blackbody from our model. For
comparison, a set of MCD SEDs  for IMBHs accreting at 1\% of critical
rate is shown. To explain the high EUV luminosity and roughly flat SED
in the EUV region, a rather high IMBH
mass is needed, $M \gtrsim 10^4 $\Msun. 

For supercritical disk this relation breaks \cite[(Poutanen \etal,
  2006)]{poutanen}, and the outcoming radiation becomes much softer,
except for the X-rays escaping along the disk axis \cite[(Fabrika \etal, 2007)]{superkarpov}. Most part of the luminosity is
supposed to be reprocessed into EUV and UV quanta, creating the nearly-flat SED of
NGC6946~ULX1. In optical/UV range contribution of the donor star may
become significant.

 In \cite{mf16_main} we make estimates for the detectability of ULXs
 with GALEX, coming to the conclusion that at least some of them (the
 sources with lower Galactic absorption) may be bright enough targets
 even for low-resolution spectroscopy.

\section{Conclusions}

We conclude that MF16 is most likely a dense shell illuminated from
inside. This can be a certain stage of the evolution of a ULXN, when the
central source is bright and the shell itself rather compact. 
We suggest that ULXs must be luminous EUV sources as well in some
cases, and may be also luminous UV sources. 

\begin{acknowledgments}
 This work was supported by the RFBR grants NN 05-02-19710, 04-02-16349, 06-02-16855.
\end{acknowledgments}

\end{document}